# SUPER-RESONANT INTRACAVITY COHERENT ABSORPTION

P. Malara[1*], C.E. Campanella[2], A. Giorgini[1], S. Avino[1], P. De Natale[3] and G. Gagliardi[1]

[1] Consiglio Nazionale delle Ricerche, Istituto Nazionale di Ottica (INO), via Campi Flegrei, 34, Comprensorio A. Olivetti, Pozzuoli, (NA), Italy.
[2] QOpSys s.r.l.s. , via Matteotti, 23, Gioia del Colle, Bari, Italy.
[3] Consiglio Nazionale delle Ricerche, Istituto Nazionale di Ottica (INO), Largo E. Fermi, 6, Firenze, Italy
*Corresponding author: pietro.malara@ino.it

**The capability of optical resonators to extend the effective radiation-matter interaction length originates from a multipass effect, hence is intrinsically limited by the resonator's quality factor. Here, we show that this constraint can be overcome by combining the concepts of resonant interaction and coherent perfect absorption (CPA). We demonstrate and investigate super-resonant coherent absorption in a coupled Fabry-Perot (FP)/ring cavity structure. At the FP resonant wavelengths, the described phenomenon gives rise to split modes with a nearly-transparent peak and a peak whose transmission is exceptionally sensitive to the intracavity loss. For small losses, the effective interaction pathlength of these modes is proportional respectively to the ratio and the product of the individual finesse coefficients of the two resonators. The results presented extend the conventional definition of resonant absorption and point to a way of circumventing the technological limitations of ultrahigh-quality resonators in spectroscopy and optical sensing schemes.**

## INTRODUCTION

The capability of an optical cavity to sustain the resonant field and store it within its volume for a finite amount of time is foundational for a vast number of physical disciplines, from classical laser spectroscopy to nonlinear optics, metrology, and obviously, lasers [1-6]. A resonant cavity allows in fact to increase the radiation-matter interaction probability, to build up the optical power and to impose a well-defined mode structure on the electromagnetic field. Because all these features scale with the finesse coefficient ($\mathcal{F}$) or the quality parameter ($Q$) of the resonator, in the last 30 years an enormous technological effort has been deployed to improve the fabrication techniques of ultra-reflective mirrors, high-Q microrings and whispering-gallery mode resonators made with the most diverse geometries and materials [7-10].

Recently, the possibility of controlling interferometrically the intracavity radiation-matter interaction was also pointed out. When a Fabry-Perot (FP) cavity is resonantly coupled with two coherent optical fields from opposite directions, an interference pattern is generated that concentrates the overall intensity inside or outside the resonator, depending on the relative phase of the input beams. For an antisymmetric phase relation and a specific value of the internal FP loss, the intensity outside the resonator can be completely extinguished, so that all the electromagnetic energy fed to the system from the two sides is trapped indefinitely and forced to dissipate within the FP. Since its first demonstration in 2010, the concept of coherent perfect absorber [11-13] (CPA) has been applied to maximize radiation coupling in various optical systems [14-18], and very recently, it has also proved an excellent scheme for sensing applications [19]. Indeed, since the perfect extinction of the input field stems from a very critical balance of interference and dissipation, the transmission of a coherent absorber when approaching the CPA condition is very sensitive to small variations of the intracavity loss.

In this work, we show that the CPA features can be combined with the resonant enhancement of a conventional ring cavity to achieve an exceptional sensitivity to the intracavity loss. The integration of these two concepts is realized in a coupled-resonator structure where a FP with a weakly absorbing medium is enclosed within the intracavity path of a larger ring resonator. Upon injection with one input field only, the system supports bidirectional super-modes split in two peaks. At the wavelengths where the both the coupled system and the FP alone are resonant, the system behaves as a super-resonant CPA, where the input fields of the internal FP are themselves the resonant modes of the coupled system. The two counterpropagating fields associated with a split mode interfere constructively/destructively inside the internal bsorber, giving rise to a "bright", nearly-transparent resonance and a "dark" resonance with exceptionally strong absorption features.

In the following sections, the transmission of such a super-resonant CPA (RCPA) cavity as a function of the internal FP loss is investigated both theoretically and experimentally. In particular, we show that in the small-loss region, the effective absorption pathlength of a dark mode is enhanced by a factor proportional to the product of the FP and ring-resonator finesse coefficients. On the contrary, the effective interaction of a bright mode depends on the ratio of the two finesse coefficients and can be therefore much smaller than physical length of the absorber. The possibility of manipulating the intracavity interaction by multiplying/dividing the enhancement factors of two resonators is of striking importance for all spectroscopy and sensing applications, for it allows to attain a detection sensitivity largely superior to that of any conventional CPA-based[19] or cavity-enhanced[20] technique, without need of ultra-high Q resonators.

## Results and discussion

The super-resonant coherent perfect absorber here demonstrated is analogous to a mode-coupled add-drop ring resonator. In a ring cavity with backscattering, the clockwise and anticlockwise modes are coupled, and a splitting of the resonant frequencies occurs, which is proportional to the amount of intracavity backreflection. The left and right wings of these split modes are commonly labelled as "symmetric" and "antisymmetric" resonances. These modes can be excited individually by injecting two fields with symmetric/antisymmetric relative phase at the two input ports of the resonator, or simultaneously, upon injection from one port only. In the latter case, for a ring with intracavity reflection $r$ and transmission $t$, the output field writes [21]:

$$\frac{E_{out}}{E_{in}} = -\frac{1}{2}\left(\frac{k^2(t+r)e^{i\frac{\beta L}{2}}}{1-\tau^2(t+r)e^{i\beta L}} + \frac{k^2(t-r)e^{i\frac{\beta L}{2}}}{1-\tau^2(t-r)e^{i\beta L}}\right) \qquad (1)$$

In equation (1), $k$ and $\tau$ are the coupling coefficients of the add/drop optical couplers (assumed identical), $\beta = 2\pi/\lambda$ is the radiation wavenumber, $L$ is the optical length of the ring and the input/output port configuration is the same of fig. 1a.

The optical system here presented (sketched in fig.1.a) can be regarded as a particular case of a coupled-mode ring resonator where the intracavity coupling element is a FP cavity of optical length $l<<L$. In this case, the reflection and transmission coefficients $r_{FP}$ and $t_{FP}$ that describe the internal backscattering are complex, wavelength-dependent functions, so that the terms $(t \pm r)$ in eq. (1) become $|t_{FP} + r_{FP}|e^{i\arg(t_{FP}+r_{FP})}$ (detailed expressions are reported in supplementary materials, eq. S1)

In such a FP-ring coupled cavity, split-resonances arise again from the symmetric and antisymmetric modes, but near the FP resonant wavelengths, where the backreflection drops abruptly, the splitting reduces and the two peaks tend to superimpose, generating a narrow interference in the transmission spectrum[22, 23], as shown in fig.1a.

In the following, we analyze the transmission of a split-mode sufficiently close to a FP resonance, but whose peaks are separated enough so that this interference can be neglected. With this simplification, the symmetric and antisymmetric fields can be regarded as the output of two independent, virtual ring resonators with effective lengths $L_{eff}^{\pm} = L + \arg(t_{FP} \pm r_{FP})/\beta$ and effective per-pass transmission $d_{eff}^{\pm} = |t_{FP} \pm r_{FP}|$ respectively.

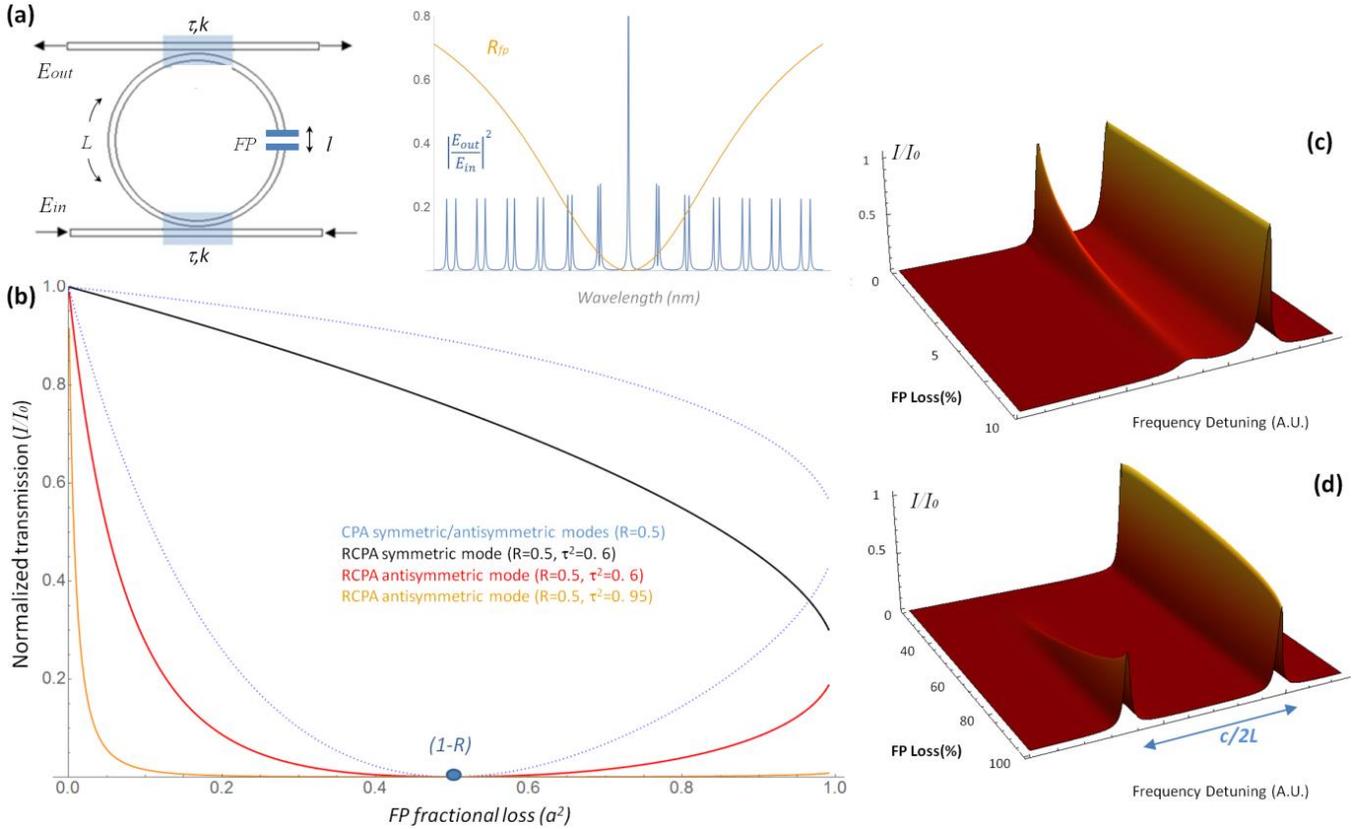

**Fig.1: (a)** RCPA resonator scheme and its transmission spectrum in the vicinity of a FP resonance. The trend of the FP reflectivity $R_{fp}$ is also plotted (orange line); **(b)** Normalized resonant transmission of a bright/dark peak pair (black and red line) as a function of the FP fractional loss (simulation parameters: R=0.5, $\tau^2$=0.8). In the same illustration, the transmission of the equivalent CPA (same FP but no ring coupling) and of the dark mode of a RCPA with higher ring transmission ($\tau^2 = 0.95$) are plotted for comparison (blue dotted line and orange line respectively). **(c, d)** Transmission of a RCPA supermode as a function of FP loss in the small and large-loss regions. In this simulation $R = 0.9$, $\tau^2 = 0.8$ (values chosen for ease of visualization).

In fig.1b, the normalized transmitted intensity $I/I_0$ of the two peaks is plotted as a function of the fractional FP power loss $a^2$. The symmetric mode (black line) is attenuated slowly but monotonically, while the antisymmetric mode (red line) exhibits a clear coherent-absorber behavior, with a rapid attenuation in the small-loss region followed by total extinction and loss-induced transparency. For this reason, we refer to these resonances as "bright" and "dark" respectively. The described transmission behavior is determined by the terms $|t_{FP} + r_{FP}|$ and $|t_{FP} - r_{FP}|$, which govern the effective loss of the symmetric/antysimmetric equivalent ring resonators. Sufficiently close to a FP resonance, $r_{FP}$ and $t_{FP}$ can be considered real and written as:

$$r_{FP} = \sqrt{R} - \frac{\sqrt{R}T(1-a^2)}{1-R(1-a^2)} \quad , \quad t_{FP} = \frac{T(1-a^2)}{1-R(1-a^2)} \tag{2}$$

In the above expressions, identical lossless FP mirrors of reflectivity $R$, transmittivity $T=1-R$ and intracavity per-pass attenuation $d^2=1-a^2$ are considered.

Any $a^2>0$, makes $r_{FP} \neq 0$ and causes resonances to split into symmetric/antisymmetric pairs. As the losses increase, a critical value $a^2 \approx 1-R$ is rapidly approached. In this point $|t_{FP} - r_{FP}| = 0$, which means the antisymmetric resonance is totally extinguished (see fig.1b). This critical point coincides with the perfect coherent absorption condition of the FP absorber.

By further increasing the FP losses, $t_{FP}$ keeps reducing. Eventually, no radiation is transmitted across the ring and the coupled system turns into a standing-wave resonator of length $L$. An evenly intense, evenly spaced mode spectrum is in fact approached through a loss-induced transparency behavior, where the transmission of the dark modes increases and eventually matches that of the bright modes. At the same time, the mode splitting approaches its maximum allowed value of $c/2L$ [22]. The transmission of a RCPA super-mode in the regions at the left and right of the critical point is illustrated in figs.1c, 1d.

To compare the described transmission behavior to that of a conventional coherent absorber, in fig.1b the transmission of the equivalent, uncoupled CPA system (same FP parameters but without external ring) is also plotted with a dotted blue line. We note that in the proposed configuration CPA is realized for a much broader range of FP loss compared to a conventional absorber. By increasing the ring transmittivity $\tau^2$, this interval can be indeed broadened to the point that the input radiation is totally extinguished for almost any intracavity loss (orange curve in fig.1a, with $\tau^2 = 0.95$).

As we also note in fig.1b, in the small-loss region prior to the perfect absorption interval, the sensitivity of the RCPA mode transmission to a variation of the FP loss is much higher than in any conventional CPA scheme, indicating an exceptional sensing capability. For a more quantitative idea, the sensitivity of the transmission to small FP losses is formalized in the following as an effective absorption pathlength $l_{eff}$.

For either the bright or the dark mode, $l_{eff}$ can be calculated as the conventional cavity enhancement[12] of the corresponding virtual resonator, whose effective per-pass power loss is $(a_{eff}^\pm)^2 = 1 - |t_{FP} \pm r_{FP}|^2$. A detailed derivation of $l_{eff}$ is reported in the supplementary materials in the assumption that the single-pass FP absorption is negligible compared both to the FP and to the ring resonator losses $\left(\frac{\alpha l}{1-R} \sim 0 \text{ and } \frac{\alpha l}{1-\Gamma} \sim 0\right)$. In this weak absorption limit, it gives:

$$l_{eff}^{dark} \cong \frac{4}{\pi^2} \mathcal{F}_{ring} \mathcal{F}_{FP} l \quad , \quad l_{eff}^{bright} \cong \frac{1}{4} \frac{\mathcal{F}_{ring}}{\mathcal{F}_{FP}} l \tag{3}$$

where $\mathcal{F}_{ring}$ and $\mathcal{F}_{FP}$, are the finesse coefficients of the ring and the FP resonator as if they were uncoupled, and $l$ is the single-pass absorption length of the FP. Eqs. (3) state that the absorption of the dark mode is enhanced by a factor equal to the product of the enhancement factors $\frac{\mathcal{F}_{ring}}{2\pi}$ and $\frac{2\mathcal{F}_{FP}}{\pi}$ of the two resonators[8,20], and can therefore exceed by far any conventional cavity-enhanced absorption. On the other hand, the pathlength enhancement of the bright mode scales with the ratio of the two resonators finesse coefficients, and when $\mathcal{F}_{ring} < 4\mathcal{F}_{FP}$ becomes smaller than unity, it results in an effective reduction of the absorption below the single-pass interaction (see supplementary materials, fig.S2).

The transmission features of the RCPA modes were observed experimentally with the setup sketched in fig.2b, built by splicing a 4cm fiber Bragg grating (FBG) FP fiber cavity in a 6m-long fiber loop (additional details are reported in the figure caption). In this setup, FP losses are conveniently introduced by bending the fiber in the region between the two FBGs, while the transmission $\tau^2$ of the external ring is tuned between 0.1 and 0.95 by means of a variable evanescent fiber coupler. The finesse of the internal FP is set by temperature-tuning the reflectivity of the FBGs.

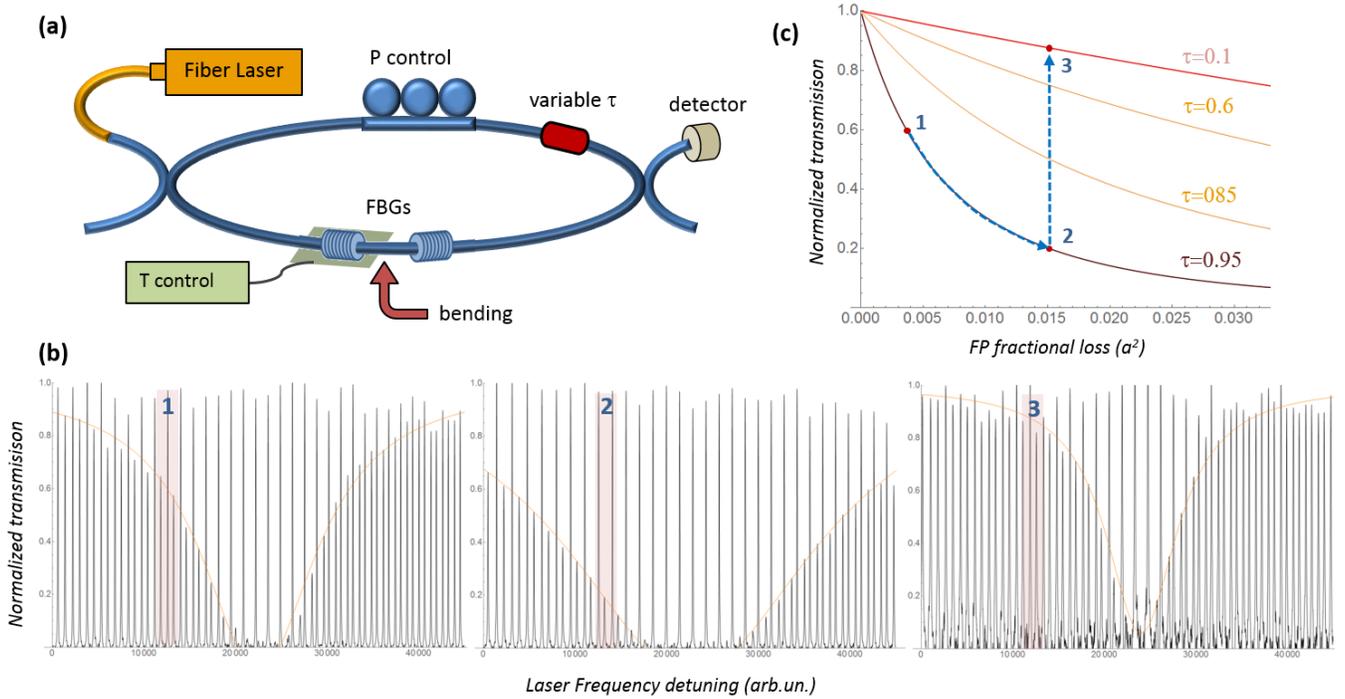

**Fig.2 (a)** RCPA experimental setup. A FBG-FP resonator is spliced within a fiber loop along with two evanescent couplers (coupling ratio 0.1%). The cavity is also equipped with a Lefevre rings device and an additional variable coupler to control the ring transmission $\tau^2$ and the intracavity polarization and injected with a narrow-linewidth fiber laser operating around 1560 nm. **(b)** Three resonant spectra recorded by scanning the laser wavelength across a resonance of the internal FP for 3 different values of FP loss and $\tau^2$. The envelope of the dark modes is highlighted in orange. **(c)** The transmission of dark modes highlighted in red in the experimental spectra shown as a trajectory on the simulated RCPA transmission.

Three examples of RCPA transmission spectra, obtained by scanning a narrow-linewidth laser across a resonance of the internal FP, are shown in fig.2c. In spectrum #1, $R = \tau^2 = 0.95$. The bright, unperturbed modes and the dark modes are clearly visible. The attenuation of the dark modes increases towards the center of the FP resonance because in that region the effective FP loss is more resonant and thus much stronger. In fact, the central modes can be thought as "ahead" in the transmission curve (closer to the perfect absorption region) while the lateral modes are behind, in the region where both the transmission and the sensitivity to loss variations are larger. It is worth noting that in spectrum #1 the only FP loss is represented by the injection loss of the FBGs, and it is already sufficient to extinguish the dark modes in the central region.

From the described configuration, spectra #2 and #3 were obtained by first increasing the FP losses $a^2$ at constant $\tau^2$, and then reducing the ring transmission $\tau^2$ to the minimum (~0.1) while keeping the loss unchanged. As expected, the first step reduces the transmission of the dark modes (note that, as expected the reduction is larger in the lateral region of the spectrum). Instead, when $\tau^2$ is reduced, the effective absorption of the dark modes becomes extremely small, even smaller than in spectrum #1, demonstrating, as predicted by eqs. 3, that the effective RCPA absorption is strongly dependent on the finesse of the external ring. For the modes highlighted in red in fig.2b, the measurement sequence is represented as a trajectory in the transmission-loss plot in fig.2c.

To investigate more quantitatively the RCPA absorption, the transmission of the dark modes was directly compared with the resonant transmission of a traditional Fabry-Perot cavity, with the very same parameters but with no coupling loop.

For this task, after recording the RCPA spectrum ($R = 0.6, \tau^2 = 0.95$) the external loop was opened by removing the variable coupler, to record the resonant transmission of the corresponding uncoupled Fabry-Perot. The measurement was repeated for 4 increasing values $\alpha_i$ of FP loss. The experimental FP and RCPA spectra (reported in fig.S3 of the supplementary materials) were fitted with a Lorentzian and an inverse Lorentzian respectively. The fitting curves, normalized to the transmission level of the bright modes $I_0$, are plotted in fig.3a.

The absorbances $\frac{\Delta I_{FP}}{I_0}$ and $\frac{\Delta I_{RCPA}}{I_0}$, measured in the central and lateral region of the fitted spectra are plotted as a function of the single-pass loss $\alpha$ in figs 3b, 3c. The values $\alpha_i$ on the x axes are calculated from the resonant FP spectra as $\alpha_i = \left(\frac{2\mathcal{F}_{FP}}{\pi}l\right)^{-1}\Delta I_{FP}$ (with $\mathcal{F}_{FP} = 6$ and $l = 4cm$). Since $\frac{\Delta I}{I_0} = \alpha l_{eff}$, the slope of the absorbance plots is a direct measurement of the effective intracavity interaction pathlength.

In the central region of the spectrum (fig. 3b) the absorbances have roughly the same slope. Again, the large intrinsic FP loss of our setup makes the central region of the RCPA spectrum fall near the perfect extinction region, where the effective extinction is high but the sensitivity to the variations of the FP loss is very small. In fact, the sensitivity of the dark modes is much larger in the lateral, off-resonance region of the spectrum (fig. 3c). Here, even though only the smaller loss values fit well on a line, a comparison with the standalone FP absorbance gives $l_{eff}^{RCPA} \sim 10\ l_{eff}^{FP}$.

It is hard to establish whether in this measurement the maximum sensitivity region of the RCPA setup is reached (where the finesse coefficients of the two resonators are effectively multiplied). Yet, the sensitivity of the dark modes to the FP loss is already one order of magnitude larger than the conventional cavity-enhanced absorption (see section 2 of supplementary materials for additional details).

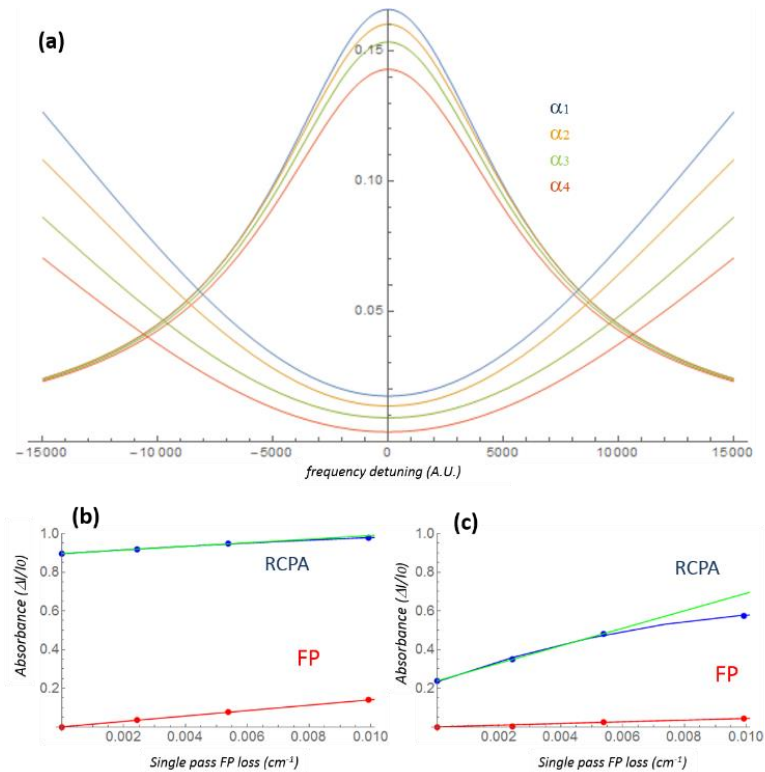

**Fig.3**. **(a)** Transmission of the uncoupled Fabry-Perot and envelope of the dark RCPA modes for 4 different values of the intracavity loss $\alpha_i$. The curves are fitted from the data reported in fig.S4 of supplementary materials along with the fit parameters. **(b,c)** absorbance $\Delta I/I_0$ of the FP and dark modes in the central and lateral regions of the spectrum.

To complete the validation of the proposed model, the high-loss regime of the RCPA transmission was also investigated. The cavity parameters were set to $R = 0.9$ and $\tau^2 = 0.8$, in order to get a more pronounced loss-induced transparency behavior (see simulated transmission curve of fig.4. The loop length was also increased, in order to keep having large number of RCPA modes within the now narrower FP resonance.

The five experimental spectra shown in fig.4, recorded for large variations of the FP loss, reproduce faithfully the model predictions. As the FP loss increase, the central dark modes enter the negative-absorption regime and start growing, while the "late" lateral modes are still being attenuated. As losses grow, the attenuation of the bright modes also starts being visible. Eventually, the RCPA turns into a standing-wave resonator, with a spectrum formed by evenly-spaced, evenly-intense peaks.

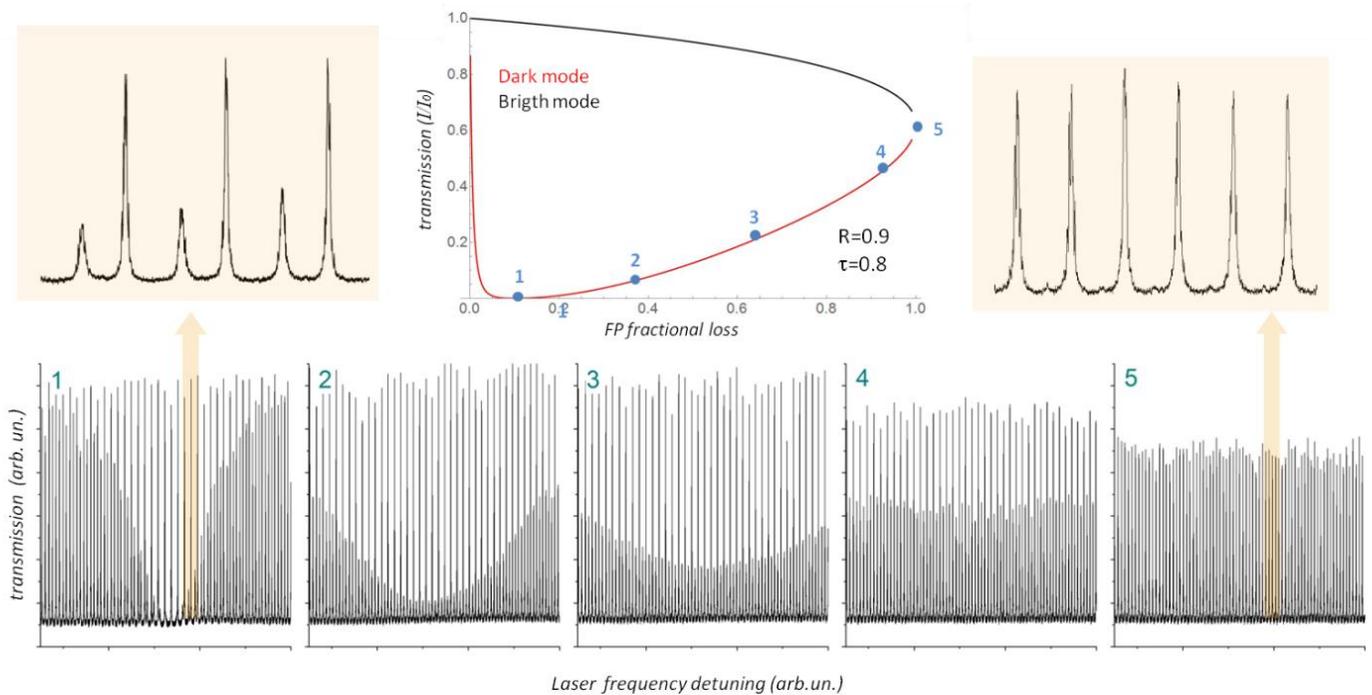

**Fig.4.** Simulated transmission curve for $R=0.9$ and $\tau^2 = 0.8$ and corresponding spectra, shown for a sequence of 5 increasing values of the FP loss. Zooms of the first and last frame are also shown to appreciate the transition from RCPA to standing-wave resonator.

In conclusion, we demonstrated and investigated a new coherent absorber configuration where the interfering fields are themselves the resonant modes of a coupled-cavity system, hence the label "super-resonant". The described effect gives rise in the spectrum of a coupled linear-ring resonator to two sets of resonances with a very different transmission: in one case nearly transparent to the FP loss (bright modes) and in the other extremely sensitive to it (dark modes). As opposed to conventional coherent absorption schemes, these features can be observed upon injection of a single resonant input field. A particularly attractive behavior of the dark resonances is pointed out: in the small-loss region prior to the CPA point their effective absorption pathlength scales with the product of the individual ring and FP finesse coefficients.

Although this region is not easily accessible with our fiber setup (because of the large injection losses of the FBGs mirrors), a clear demonstration of enhanced absorption sensitivity (by one order of magnitude compared to the conventional cavity output) and near-transparency of the RCPA modes was provided. With an accurately designed low-loss setup, the presented geometry may open the way to spectroscopic and optical detection schemes that surpass the sensitivity limits currently imposed by high-Q resonators fabrication.

**Additional information**

The authors declare no competing financial interests.

# Supplemental material

**In this supplemental materials we report the detailed derivations of equations 3a and 3b of the main text, describing the effective interaction pathlength of the symmetric and antisymmetric modes of a super-resonant coherent perfect absorber (RCPA). Additional experimental details and data are also provided.**

In this section we work out the sensitivity of the symmetric and antisymmetric modes of the RCPA modes to an intra-FP absorption in terms of equivalent absorption pathlength. For simplicity, we consider an internal FP with identical, lossless mirrors of reflectivity R and transmissivity T=1-R, and a refractive index $n_{fp}<n_{ring}$ (the optical field gains a π phase factor only upon reflection towards the ring). In these conditions, the fields reflected and transmitted by the FP are:

$$r_{FP}(\beta) = -\sqrt{R}\left(1 - \frac{Td^2 e^{2i\beta l}}{1 - e^{2i\beta l}Rd^2}\right), \quad t_{FP}(\beta) = \frac{Td e^{i\beta l}}{1 - e^{2i\beta l}Rd^2} \quad (S0)$$

Where $\beta$ is the optical wavenumber, *d* is the per-pass field intracavity transmission of the FP and *l* its length. In the close vicinity of a resonance, where $\beta l \sim \pi$, the above expressions become

$$r_{FP} = -\sqrt{R}\left(1 - \frac{Td^2}{1 - Rd^2}\right), \quad t_{FP} = -\frac{Td}{1 - Rd^2} \quad (S1)$$

Equation S1 can be used to calculate $|t_{FP} + r_{FP}|$ and $|t_{FP} - r_{FP}|$. After some algebraic passages, we get:

$$|t_{FP} + r_{FP}| = \left|\frac{\sqrt{R} + d}{1 + \sqrt{R}d}\right| \tag{S2a}$$

$$|t_{FP} - r_{FP}| = \left|\frac{d - \sqrt{R}}{1 - \sqrt{R}d}\right| \tag{S2b}$$

In the presence of an absorbing sample (with absorption coefficient $\alpha$) homogeneously distributed over the internal length of the FP resonator, we can consider $d = e^{-\frac{\alpha}{2}l}$. For small absorption ($\alpha l \to 0$), $e^{-\frac{\alpha}{2}l} \sim 1 - \frac{\alpha l}{2}$. Substituting $d$ in Eqs. (S2):

$$|t_{FP} + r_{FP}| = \frac{1 + \sqrt{R} - \frac{\alpha}{2}l}{1 + \sqrt{R} - \sqrt{R}\frac{\alpha}{2}l} \tag{S3a}$$

$$|t_{FP} - r_{FP}| = \frac{1 - \sqrt{R} - \frac{\alpha}{2}l}{1 - \sqrt{R} + \sqrt{R}\frac{\alpha}{2}l} \tag{S3b}$$

By dividing numerator and denominator by *1-R* we obtain:

$$|t_{FP} + r_{FP}| = \frac{\frac{1}{1-\sqrt{R}} - \frac{\alpha l}{2(1-R)}}{\frac{1}{1-\sqrt{R}} - \frac{\sqrt{R}}{2}\frac{\alpha l}{1-R}} \tag{S4a}$$

$$|t_{FP} - r_{FP}| = \frac{\frac{1}{1+\sqrt{R}} - \frac{\alpha l}{2(1-R)}}{\frac{1}{1+\sqrt{R}} + \frac{\sqrt{R}}{2}\frac{\alpha l}{1-R}} \tag{S4b}$$

At this point, we introduce the assumption that the single-pass absorption in the FP is negligible compared to the FP outcoupling: $\frac{\alpha l}{1-R} \sim 0$. In this approximation, we consider $|t_{FP} + r_{FP}|$ and $|t_{FP} - r_{FP}|$ as functions of $\frac{\alpha l}{1-R}$ and linearize them around 0, obtaining

$$|t_{FP} + r_{FP}| = 1 + \frac{1}{2}\sqrt{R}\left(\frac{1-\sqrt{R}}{1+\sqrt{R}}\right)\alpha l \tag{S5a}$$

$$|t_{FP} - r_{FP}| = 1 - \frac{1}{2}\sqrt{R}\left(\frac{1+\sqrt{R}}{1-\sqrt{R}}\right)\alpha l \tag{S5b}$$

As a final remark, we recall that the finesse of a Fabry-Perot is given by $\mathcal{F}_{FP} = \frac{\pi\sqrt{R}}{1-R}$, and that $\frac{1+\sqrt{R}}{1-\sqrt{R}} \sim 4\frac{\mathcal{F}_{FP}}{\pi}$, as shown in fig.S1. With this in mind, Eqs (S5) can be written as

$$|t_{FP} + r_{FP}| = 1 + \frac{1}{8}\sqrt{R}\left(\frac{\mathcal{F}_{FP}}{\pi}\right)^{-1}\alpha l \tag{S6a}$$

$$|t_{FP} - r_{FP}| = 1 - 2\sqrt{R}\frac{\mathcal{F}_{FP}}{\pi}\alpha l \tag{S6b}$$

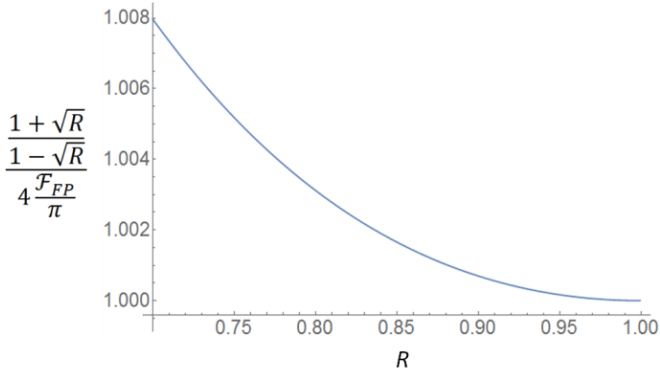

**Fig.S1**: validity of the approximation used in Eqs.(S6).

We can write the RCPA transmission by substituting $t_{FP}$ and $r_{FP}$ in equation (1) of the main text:

$$\frac{E_{out}}{E_{in}} = -\frac{1}{2}\left(\frac{k^2|t_{FP}+r_{FP}|e^{-i\left[\frac{\beta L}{2}+\arg(t_{FP}+r_{FP})\right]}}{1-\tau^2|t_{FP}+r_{FP}|e^{-i[\beta L+\arg(t_{FP}+r_{FP})]}} + \frac{k^2|t_{FP}-r_{FP}|e^{-i\left[\frac{\beta L}{2}+\arg(t_{FP}-r_{FP})\right]}}{1-\tau^2|t_{FP}-r_{FP}|e^{-i[\beta L+\arg(t_{FP}-r_{FP})]}}\right)$$

$k$ and $\tau$ are the coupling coefficients of the ring optical couplers (assumed identical). For separated peaks (no interference), the intensity of the symmetric and antisymmetric resonances, normalized to their peak transmission $I_0$ (in absence of absorption) is:

$$\frac{I_{sym}}{I_0} = \frac{K^2|t_{FP}+r_{FP}|^2}{(1-\Gamma|t_{FP}+r_{FP}|)^2} \tag{S7a}$$

$$\frac{I_{asym}}{I_0} = \frac{K^2|t_{FP}-r_{FP}|^2}{(1-\Gamma|t_{FP}-r_{FP}|)^2} \tag{S7b}$$

Where $\Gamma = \tau^2$ and $K = k^2$. We note that the above expressions are analogous to the normalized transmissions of two independent rings with per-pass field intracavity transmission coefficient $d_{eff} = |t_{FP} \pm r_{FP}|$ respectively.

We now substitute in equations (S7) the $|t_{FP} \pm r_{FP}|$ calculated in equation (S6). For ease of notation, we use a generic expressions $1 + A^{\pm}\alpha l$. We will distinguish symmetric and antisymmetric transmission at the end of the calculation, by substituting $A^{\pm}$ with the coefficients of $\alpha l$ of equation (S6).

$$\frac{I}{I_0} = \frac{K^2(1+(A^{\pm}\alpha l)^2+2A^{\pm}\alpha l)}{1+\Gamma^2+\Gamma^2(A^{\pm}\alpha l)^2-2\Gamma-2\Gamma A^{\pm}\alpha l} \tag{S8}$$

Neglecting the quadratic terms in $\alpha l$ we rearrange in

$$\frac{I_{out}}{I_0} = \frac{K^2}{(1-\Gamma)^2-2\Gamma A^{\pm}\alpha l} + \frac{2K^2 A^{\pm}\alpha l}{(1-\Gamma)^2-2\Gamma A^{\pm}\alpha l} \tag{S9}$$

Again the second term on the right-end side is negligible for $\alpha l \to 0$.

$$\frac{I_{out}}{I_0} = \left(\frac{K}{1-\Gamma}\right)^2 \frac{1}{1 - \frac{2\Gamma}{1-\Gamma} A^{\pm} \alpha l} \tag{S10}$$

Then, we get to the absorbance:

$$\frac{\Delta I^{\pm}}{I_0} = 1 - \frac{1}{1 - 2\Gamma A^{\pm} \frac{\alpha l}{1-\Gamma}} \tag{S11}$$

Analogously to what done previously for the internal FP cavity, if the single-pass absorption is small also compared to the coupling losses of the ring resonator $\left(\frac{\alpha l}{1-\Gamma} \sim 0\right)$, we can linearize the absorbance (S11) obtaining

$$\frac{\Delta I^{\pm}}{I_0} = 2\Gamma A^{\pm} \frac{\alpha l}{1-\Gamma} \tag{S12}$$

At this point we can finally write the absorbance of the symmetric and antisymmetric modes by substituting the coefficients $A^{\pm}$ with the coefficient of equation (S6):

$$\frac{\Delta I_{sym}}{I_0} = \sqrt{\Gamma R} \frac{1}{4} \frac{\mathcal{F}_{ring}}{\mathcal{F}_{FP}} \alpha l \tag{S13a}$$

$$\frac{\Delta I_{asym}}{I_0} = \sqrt{\Gamma R} \frac{4}{\pi^2} \mathcal{F}_{ring} \mathcal{F}_{FP} \alpha l \tag{S13b}$$

Now, considering that the absorbance is defined as $\frac{\Delta I}{I_0} = \alpha l_{eff}$, we get the following expression for the effective pathlength of the symmetric and the antisymmetric modes:

$$l_{eff}^{sym} = \sqrt{\Gamma R} \frac{1}{4} \frac{\mathcal{F}_{ring}}{\mathcal{F}_{FP}} l \tag{S14a}$$

$$l_{eff}^{asym} = \sqrt{\Gamma R} \frac{4}{\pi^2} \mathcal{F}_{ring} \mathcal{F}_{FP} l \tag{S14b}$$

Eqs.(S14) (corresponding to Eq.(2) of the main text) state that the pathlength enhancement in the antisymmetric mode scales with the product of the ring and the FP enhancement factors, and is therefore much larger than the traditional resonant enhancement. For the symmetric mode instead, the pathlength enhancement scales as the ratio of the ring and FP enhancement factors, and can therefore be even smaller than 1, which means that the effective absorption of the symmetric mode is smaller than the single-pass absorption.

The situation is illustrated in fig. S2. In the top layer, the modal absorption of a RCFP and the equivalent FPs are plotted for different values of the FP intracavity loss. In the bottom layer, the symmetric and antisymmetric mode absorbance is plotted for a fixed $\mathcal{F}_{FP}$ and different values of $\mathcal{F}_{ring}$, along with the absorbance of the uncoupled FP mode and the single-pass absorbance (dashed black lines).

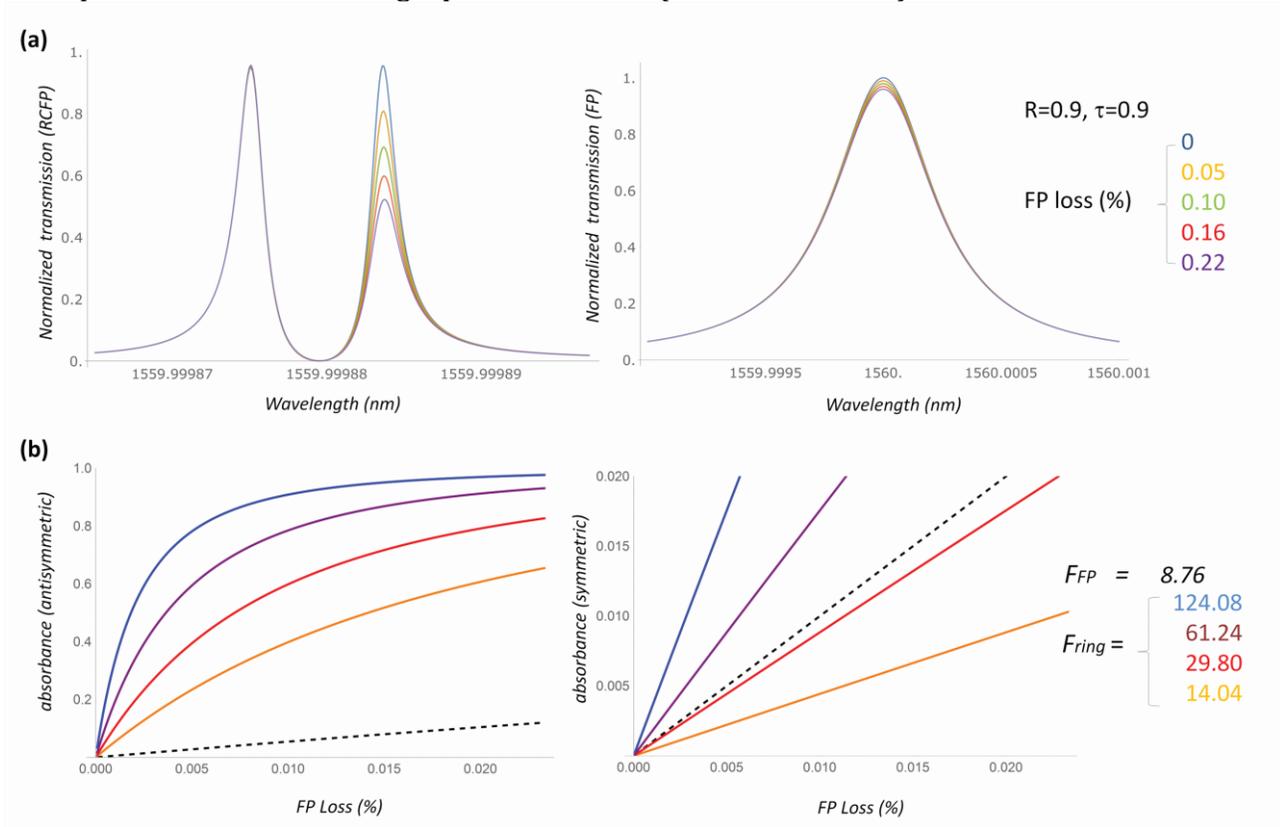

**Fig.S2** – weak-absorption behavior of the RCPA modes. **(a)** Resonant modes of the RCPA (left) and the internal FP alone (without ring-coupling) compared for different values of the intracavity loss. **(b)** Absorbance curves of the RCPA dark (left) and bright (right) modes calculated for different values of the ring resonator finesse. The absorbance of the internal FP alone is plotted in dotted line for comparison.

2 – Absorbance measurements.

In these measurements, in an attempt to reduce the saturation of the antisymmetric absorption at the center of the spectrum, the finesse of the internal FP resonator was also reduced from the $F_{FP}{\sim}25$ to $F_{FP}{\sim}6$ by tuning the reflectivity of the cavity FBG-mirrors at the laser wavelength to R=0.6.

For 4 different loss levels, the coupled-resonator transmission and the open-loop Fabry-Perot transmission were acquired. The recorded spectra are shown in Fig. S3. By fitting these data as described in the main text, and normalizing to the same $I_0$ level, Fig.3a of the main text was obtained. The best fit parameters are shown in table S1. In table S2 instead the best fit parameters for the absorbance curves calculated in the central and lateral region of the acquired spectra are reported.

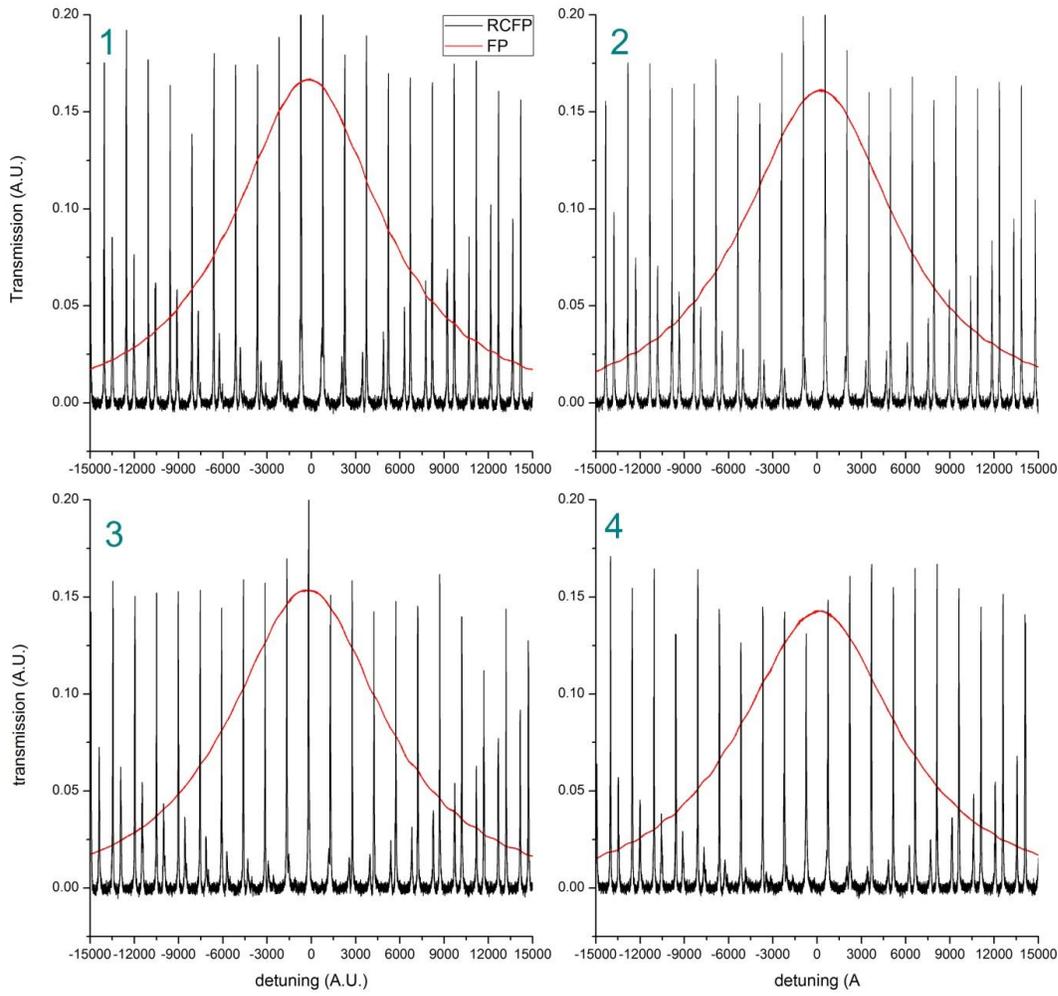

Fig.S3: FP and RCPA spectra (red and black lines), acquired by scanning a narrow-linewidth laser (operating wavelength ~1560nm) across a resonance of the internal FP. The four pairs of spectra shown correspond to 4 different levels of FP losses.

| FP resonance: $y = y_0 + \frac{2A}{\pi}\frac{w}{4x^2+w^2}$ | | | |
|---|---|---|---|
|  | y0 | w | A |
| 1 | -0.142 | 11981.459 | 76569.234 |
| 2 | -0.149 | 12421.831 | 77028.467 |
| 3 | -0.155 | 12561.539 | 74539.256 |
| 4 | -0.156 | 12937.475 | 71476.272 |
| RCPA dark modes envelope: $y = y_0 - A\frac{w}{x^2+w^2}$ | | | |
|  | y0 | w | A |
| 1 | 0.416 | 24429.088 | 9742.841 |
| 2 | 0.323 | 22608.871 | 6998.455 |
| 3 | 0.361 | 28346.453 | 9989.890 |
| 4 | 0.313 | 28593.335 | 8859.598 |

Table S1: Curve models and best-fit paramenters for the acquired FP resonances and the envelope of the RCPA dark resonances (best fit curves in fig 3a of the main text).

| Slope of the absorbance curve (from linear fit) | | |
|---|---|---|
| Cavity | FP | RCPA |
| Central | 12.732 | 8.497 |
| Lateral (first 3 points) | 3.961 | 41.153 |

Table S2: linear-fit slopes for the absorbance measurements plotted in Fig.3b,3c of the main text.